\documentclass[twocolumn,amsmath,amssymb,amsfonts,superscriptaddress,floatfix,showpacs]{revtex4-1}
\usepackage[english]{babel}
\usepackage[colorlinks=true,citecolor=green,linkcolor=red,urlcolor=blue]{hyperref}
\usepackage{braket}
\usepackage{bbm}
\usepackage{graphicx}
\usepackage{color}
\usepackage{algorithm}
\usepackage{algpseudocode}

\newcommand{\tr}{\operatorname{Tr}}
\newcommand{\ttr}{\operatorname{tTr}}

\newcommand{\ec}{\ensuremath{\mathrm{e}}}

\newcommand{\Fig}[1]{Fig.~\ref{#1}}

\newcommand{\nrm}[1]{\ensuremath{\left\Vert #1 \right\Vert}}

\newcommand{\Eq}[1]{Eq.~\eqref{#1}}
\newcommand{\skippart}[1]{}
\newcommand{\simplesection}[1]{\emph{#1} ---}

\begin{document}

\title{Renormalization group flows of Hamiltonians using tensor networks}

\author{M.~\surname{Bal}}
\affiliation{Department of Physics and Astronomy, Ghent University, Krijgslaan 281, S9, B-9000 Ghent, Belgium}
\author{M.~\surname{Mari\"en}}
\affiliation{Department of Physics and Astronomy, Ghent University, Krijgslaan 281, S9, B-9000 Ghent, Belgium}
\author{J.~\surname{Haegeman}}
\affiliation{Department of Physics and Astronomy, Ghent University, Krijgslaan 281, S9, B-9000 Ghent, Belgium}
\author{F.~\surname{Verstraete}}
\affiliation{Department of Physics and Astronomy, Ghent University, Krijgslaan 281, S9, B-9000 Ghent, Belgium}
\affiliation{Vienna Center for Quantum Technology, University of Vienna, Boltzmanngasse 5, 1090 Vienna, Austria}

\begin{abstract}
A renormalization group flow of Hamiltonians for two-dimensional classical partition functions is constructed using tensor networks. Similar to tensor network renormalization ([G.~Evenbly and G.~Vidal, Phys.~Rev.~Lett.~115,~180405 (2015)], [S.~Yang, Z.-C.~Gu, and X.-G~Wen, Phys.~Rev.~Lett.~118, 110504 (2017)]) we obtain approximate fixed point tensor networks at criticality. Our formalism however preserves positivity of the tensors at every step and hence yields an interpretation in terms of Hamiltonian flows. We emphasize that the key difference between tensor network approaches and Kadanoff's spin blocking method can be understood in terms of a change of local basis at every decimation step, a property which is crucial to overcome the area law of mutual information. We derive algebraic relations for fixed point tensors, calculate critical exponents, and benchmark our method on the Ising model and the six-vertex model.
\end{abstract}
\maketitle

\simplesection{Introduction} The study of phase transitions and critical properties of lattice models has long been at the center of statistical physics. Universal properties of critical systems can be captured by conformal field theories (CFTs), which act as low-energy effective descriptions of critical models, and whose scaling dimensions can be related to the critical exponents of asymptotic correlation functions. One way to gain insight into these phenomena is through real-space renormalization group (RG) methods, which predate the development of the modern renormalization group and can be traced back to Kadanoff's block spin procedure \cite{Kadanoff1966,*Kadanoff1975,*Kadanoff1976}. In his treatment of block spin methods on the lattice, Wilson emphasized that one should be able to do precise numerical calculations using pure RG methods combined with approximations based only on locality \cite{Wilson1975}. For real-space RG to work, the effective Hamiltonian at every step should be dominated by short-range interactions as interactions of arbitrary complexity are generated in subsequent iterations. Additionally, the calculation of any particular term in the coarse-grained Hamiltonian should involve but a small number of fine-grained spins.

Tensor networks are efficient, local, real-space variational ans\"atze for many-body wavefunctions, which are constructed by mimicking the spatial distribution of entanglement and correlations. Renormalization group methods based on tensor networks satisfy Wilson's requirements insofar as their inherent real-space locality and finite bond dimension restrict the range of newly generated effective interactions and provide a controlled approximation that can be systematically improved.

For two-dimensional lattice systems, the tensor renormalization group (TRG) algorithm \cite{Levin2007,Gu2008} put the idea of tensor network renormalization (TNR) into practice in a most explicit way. Wholly based on truncation using singular value decomposition (SVD), this algorithm works extremely well for gapped systems because of the same entanglement reasons that explain the success of the density matrix renormalization group (DMRG) for quantum spin chains \cite{Levin2007}. Despite remarkable accuracy in determining critical exponents for finite systems, none of the methods based on TRG \cite{Xie2009,Zhao2010,Xie2012} is sustainable in the sense that it is capable of yielding true (approximate) fixed points tensors at criticality \cite{Evenbly2015b}. Recently, novel TNR algorithms respectively based on the multi-scale entanglement renormalization ansatz (MERA-TNR) \cite{Evenbly2015b,Evenbly2015,Evenbly2015a,Evenbly2016} and matrix product states (Loop-TNR) \cite{Yang2015} have been developed which do manage to flow to approximate fixed point tensors, even at criticality. Our work has been inspired by the latter proposal which formulates TNR in terms of periodic matrix product states (MPS). For the 2D classical Ising model, impressive numerical results have been obtained that seem to defy the breakdown of TRG at criticality.

In this paper, we demonstrate how tensor networks can be used to achieve explicit real-space RG flows in the space of classical Hamiltonians. To this end, we have developed a sustainable and manifestly nonnegative TNR method (TNR$_{+}$) to coarse-grain classical partition functions. By virtue of the element-wise nonnegativity of all tensors involved, we can explicitly associate a Hamiltonian to the fixed point tensors of the RG flow generated by our algorithm. We thus believe our work opens up the possibility to begin to address one of the main concerns raised by the traditional real-space RG community about all TNR schemes: the lack of an insightful RG interpretation of what are essentially supposed to be real-space RG methods \footnote{``[...] the more recent tensor-style work often employs indices which are summed over hundreds of values, each representing a sum of configurations of multiple spinlike variables. All these indices are generated and picked by the computer. The analyst does not and cannot keep track of the meaning of all these variables. Therefore, even if a fixed point were generated, it would not be very meaningful to the analyst. In fact, the literature does not seem to contain much information about the values and consequences of fixed points for the new style of renormalization'' \cite{Efrati2014}}.

\simplesection{Tensor network renormalization} The salient features shared by all TNR algorithms developed up to now are twofold. First, the breaking apart of the tensor product structure, which was introduced in TRG by splitting tensors using SVD, is crucial to the construction of new effective degrees of freedom and the removal of correlations. The reason why Kadanoff's spin blocking fails can be traced back to the bounds on correlations imposed by the mutual information between a block and its environment. In order to overcome this barrier, it is essential to reorganize degrees of freedom by doing a local basis transformation. Secondly, both MERA-TNR and Loop-TNR address the additional need to extend the domain of the coarse-graining step to act on a block of sites in order to remove intra-block correlations. The disentangling power of both MERA-TNR and Loop-TNR can be found in surrounding a block of sites with a coarse-graining operator. This explains, for instance, why there is no way for TRG, which acts locally on each site, to detect the short-range correlations that it sets out to remove at criticality.

\simplesection{Coarse-graining nonnegative tensor networks}
Consider a two-dimensional bipartite square lattice of $N$ classical spins $\{s_{i}\}$ described by an energy functional $H(s_{1},s_{2},\ldots)$. The classical statistical partition function is then given by
\begin{align}
\mathcal{Z}=\ec^{\beta F}=\tr_{\{ s_{1}, s_{2},\ldots \}} \ec^{\beta H(s_{1},s_{2},\ldots)},
\end{align}
where $F=E-TS$ denotes the free energy. If we imagine the spins living on the vertices of the lattice, the Boltzmann weight of a site depends on the configuration of the bonds connected to the site. We can write these probabilities as a rank-four tensor $A_{ijkl}$, so that the sum over all configurations in the partition function boils down to contracting a nonnegative tensor network,
\begin{align}
	\mathcal{Z}[A]=\ttr \bigotimes A_{ijkl}. \label{nonnegtn}
\end{align}
By coarse-graining tensor networks, we then refer to a real-space RG procedure constructing a sequence of partition functions $\mathcal{Z}[A^{0}] \to \mathcal{Z}[A^{1}] \to \ldots \to \mathcal{Z}[A^{s}]$, where each effective partition function is defined on a coarser lattice than the one before, until we are left with a single effective site after $s \approx \log_{2}(N)$ iterations. If we now want to additionally retain element-wise nonnegativity of all involved tensors at every step, we cannot resort to using SVD, which is the backbone of all other TNR approaches. Instead, we are led to nonnegative matrix factorization (NMF) algorithms \cite{suppmat} to approximate the following matrix factorization problem: given an element-wise nonnegative matrix $A \in \mathbb{R}_{+}^{m \times n}$ and a rank $k \leq \min (m,n)$, find the matrices $X \in \mathbb{R}_{+}^{m \times k}$ and $Y \in \mathbb{R}_{+}^{k \times n}$ minimizing $\nrm{A-XY}^2_{F}$ \footnote{Given the nature of the problem, one might expect an $l_{1}$-norm. In practice, tackling the $l_{1}$-norm optimization problem is not economical due to the large number of constraints, hence the relaxation to a smooth optimization in practice.}.

\begin{figure}[t]
 \includegraphics[width=0.85\linewidth,keepaspectratio=true]{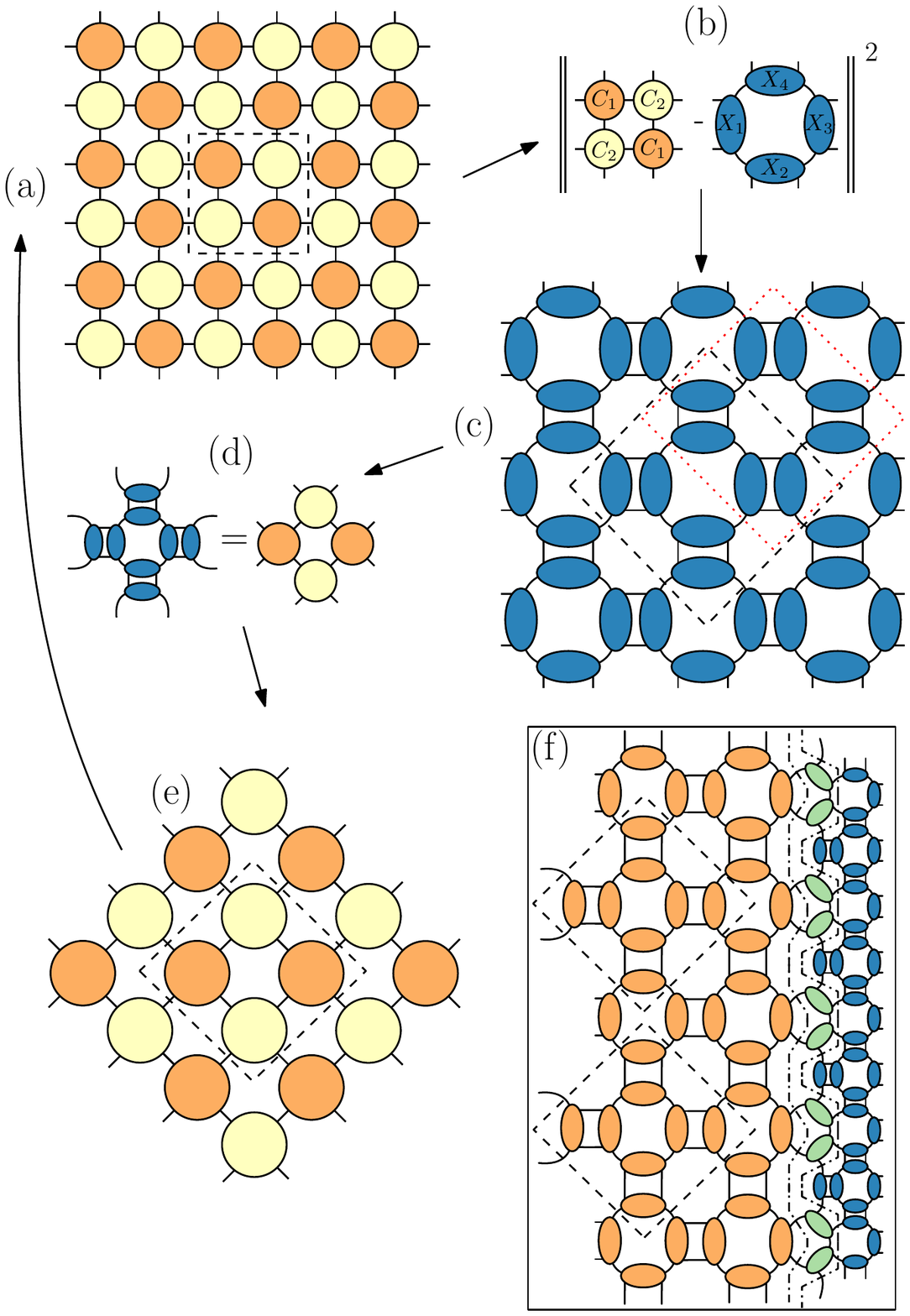}
 \caption{(a-e) Illustration of a single step of the TNR$_{+}$ algorithm. (a) Starting from a bipartite square lattice, (b) we approximate the periodic MPS representing a block of four sites by a rotated version (c) with a different tensor product structure, and (d) contract these numerically optimized tensors exactly to (e) arrive at a coarse-grained tilted lattice. (f) Iterating the TNR$_{+}$ procedure in the presence of an open boundary generates a stochastic MERA.}
 \label{fig:construction}
\end{figure}

Now let us focus on a block of four adjacent sites (\Fig{fig:construction}(a)), which we, following Yang \textit{et al.} \cite{Yang2015}, interpret as a periodic four-site matrix product state (MPS) with respective physical and virtual dimensions. The local coarse-graining procedure then proceeds according to the canonical real space RG steps by (i) \textit{introducing new effective degrees of freedom}, which here involves approximating the local block with an ansatz that has a different tensor product structure in order to remove short-range correlations (\Fig{fig:construction}(b)), (ii) \textit{summing over old degrees of freedom} by recombining the optimized tensors into new coarse-grained tensors $C_{1}$ and $C_{2}$ (\Fig{fig:construction}(d)). The virtual dimension in step (i) can be increased at will, which in turn determines the local dimension of the degrees of freedom living on the new lattice. While step (ii) explicitly sums over the old outer (physical) degrees of freedom to construct the coarse-grained tensors, step (i) also contains an implicit summation over the old inner (virtual) degrees of freedom. After a single RG step, the roles of the physical and virtual MPS dimensions have interchanged and the linear dimension of the lattice is reduced by $\sqrt{2}$. The tensors in \Fig{fig:construction}(e) then serve as input to the next step, where we take into account that we have to break up the tensor product structure again. Notice that in \Fig{fig:construction}(c) we identify the coarse-grained lattice with the ``vertex'' configuration inside the dashed bounding box and not the ``plaquette'' configuration inside the dotted one. Even though a priori they look similar, the latter configuration leads to worse numerics which can be understood by it not being able to remove short-range correlations of the corner double-line (CDL) form \cite{suppmat}.

\simplesection{Renormalization group flow}
In \Fig{fig:construction}(f) we have depicted the tensor network generated by the action of TNR$_{+}$ on an open boundary of the lattice. In much the same way as TRG produces a tree tensor network and MERA-TNR a multi-scale entanglement renormalization ansatz \cite{Evenbly2015}, our TNR$_{+}$ algorithm builds up a nonnegative tensor network approximation to the leading eigenvector of the transfer matrix. Given the nonnegativity and the alternating pattern of one iteration ``disentangling'' (blue tensors) and the next one reducing the degrees of freedom (green tensors), TNR$_{+}$ can be said to generate a stochastic MERA \footnote{Note that all these boundary tensor networks are but different low-rank tensor network approximations of the leading eigenvector of the transfer matrix written as a matrix product operator (MPO) \cite{Haegeman2016}.}. If we instead track the action of TNR$_{+}$ around an open impurity, we end up with the following MPO after two iterations \cite{suppmat},
\begin{align}
	R=\vcenter{\hbox{ \includegraphics[width=0.24\linewidth]{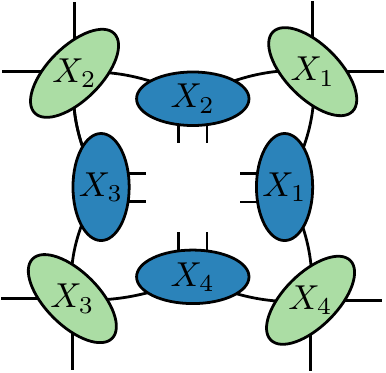} }}.\label{eq:radialmpo}
\end{align}
In the scale invariant regime of the RG flow, this MPO is identified with the radial transfer matrix \cite{Evenbly2015}, which can be diagonalized to give $R=\sum_{\alpha} 2^{-\Delta_{\alpha}} \ket{\alpha}\bra{\alpha}$. Here, the scaling dimensions $\Delta_{\alpha}$ and approximate lattice representations $\ket{\alpha}$ of the primary fields and descendants of the underlying CFT description are found only if the relative gauge freedom of the coarse-grained partition functions has been fixed, i.e.~if the degrees of freedom we deem equivalent after two iterations do indeed match \cite{suppmat}. For critical systems, we thus end up with a window of an approximately invariant alternating sequence of partition functions $\mathcal{Z}[C_{1,A}^{*},C_{2,A}^{*}] \to \mathcal{Z}[C_{1,B}^{*},C_{2,B}^{*}]$ after the initial part of the flow has sufficiently suppressed irrelevant lattice details and up until the accumulated truncation errors eventually prevail.

We can furthermore consider the fixed point equations of TNR$_{+}$ as an algebraic set of equations in their own right by finding tensors which (approximately) satisfy
\begin{align}
	\vcenter{\hbox{ \includegraphics[width=0.30\linewidth]{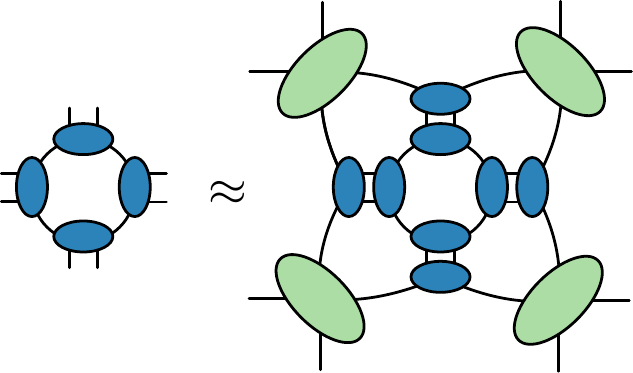} }} \mathrm{\ and\ } \vcenter{\hbox{ \includegraphics[width=0.37\linewidth]{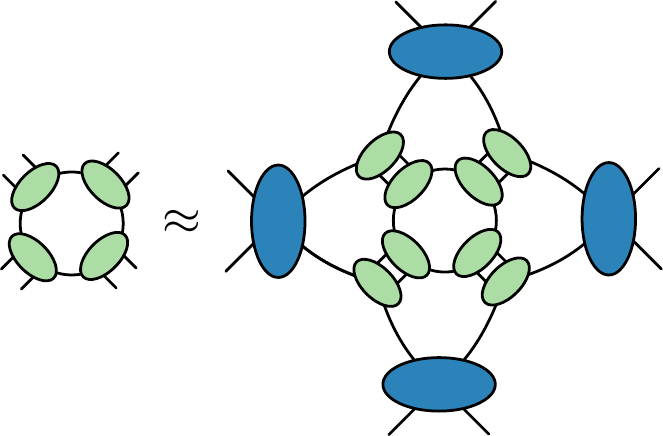} }}.\label{eq:fpeqs}
\end{align}
Exact solutions of these equations include trivial product states and GHZ states corresponding to gapped infrared fixed points, potentially with symmetry breaking. Including additional symmetry constraints, there might exist non-trivial solutions which approximately yet accurately satisfy the RG fixed point equations. The sets of these solutions and their stability under perturbations could then point towards the conditions required for a classification of all possible (approximate) RG fixed points of TNR schemes \footnote{Note that by working with symmetric tensor networks, we can also extract CFT data of non-local fields if we modify \Eq{eq:radialmpo} to include a matrix product operator (MPO) threading through the transfer matrix (which encodes the anti-periodic boundary conditions and reduces to just a string of matrices for the tensor product symmetry considered in Ref.\cite{Evenbly2016}). Similarly, the algebraic fixed point equations \Eq{eq:fpeqs} can be modified to include an additional MPO symmetry string.}\cite{balunpub}.

\simplesection{Application to classical partition functions}
We have benchmarked our algorithm on the classical Ising model and the six-vertex model. The partition function of the ferromagnetic Ising model can be encoded by associating a tensor $A_{ijkl}= \sum_{s} \left(\sqrt{a}\right)_{is} \left(\sqrt{a}\right)_{js} \left(\sqrt{a}\right)_{ks} \left(\sqrt{a}\right)_{ls}$ to each vertex, where $a_{mn}=[e^{\beta}\openone + e^{-\beta} X]_{mn}$ denotes the contribution of the interaction between spins $m$ and $n$. The Ising model exhibits a phase transition at the critical temperature $T_{c}=2/\ln(1+\sqrt{2})$ described by a free fermion $c=1/2$ CFT, separating the $Z_{2}$ symmetry breaking phase for $T<T_{c}$ from a trivial disordered phase for $T>T_{c}$. The partition function of the zero-field six-vertex model can be written in terms of the non-vanishing tensor elements $A_{1111}=A_{2222}=a$, $A_{2112}=A_{1222}=b$, and $A_{2121}=A_{1212}=c$, where $a,b,c$ denote the Boltzmann weights of the allowed bond configurations. In terms of the parameter $\Delta=(a^2+b^2-c^2)/(2ab)$, the six-vertex model has a phase boundary determined by $|\Delta|=1$ {which separates four phases: two ferroelectric phases for $\Delta>1$, an antiferroelectric phase for $\Delta<-1$, and a gapless disordered phase for $-1 < \Delta < 1$. The six-vertex model belongs to special classes of Hamiltonians which violate the universality hypothesis in that its phase diagram contains a continuum of critical points with continuously varying critical exponents captured by a free boson $c=1$ CFT. In what follows, we will consider the example of spin ice, i.e.~$a=b=c=1$ and $\Delta=0.5$.

\begin{figure}[t]
 \includegraphics[width=1.0\linewidth,keepaspectratio=true]{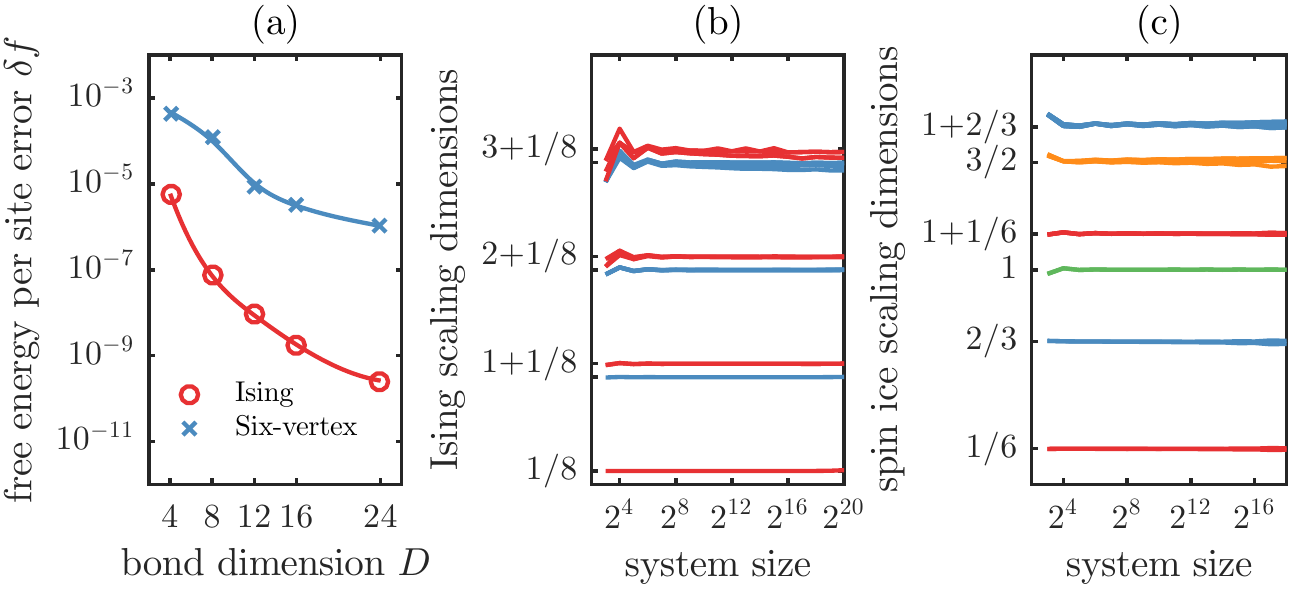}
 \caption{TNR$_{+}$ simulations for the critical Ising model and spin ice. (a) Relative error of the free energy per site in function of TNR$_{+}$ bond dimension ($N=2^{32}$ sites). (b,c) Scaling dimensions extracted from the linear transfer matrix MPO \Eq{eq:lintm42} in function of RG step (Ising $D=16$, spin ice $D=12$).}
 \label{fig:numerics}
\end{figure}

In \Fig{fig:numerics}(a), the relative error of the free energy per site $f=-\log (Z)/N$ is plotted at criticality in function of the bond dimension. We observe very accurate free energies, with the difference in accuracy between the simulations of the two models reflecting the less trivial nature of the six-vertex model. To study the implicit approximate scale invariance of the RG flow, we calculate the smallest scaling dimensions from the linear transfer matrix MPO constructed from $4 \times 2$ effective partition function tensors,
\begin{align}
	M=\vcenter{\hbox{ \includegraphics[width=0.30\linewidth]{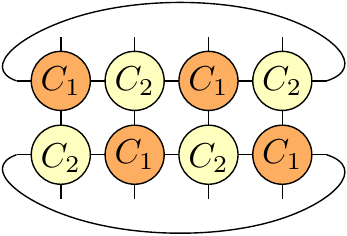} }},\label{eq:lintm42}
\end{align}
in function of system size (or, equivalently, iteration step) in \Fig{fig:numerics}(c,d) \cite{suppmat}. We observe that the numerically obtained implicit fixed point is stable under subsequent coarse-graining and remains so for a prolonged number of steps, in agreement with other TNR approaches \cite{Evenbly2015b,Yang2015}\footnote{Eventually though, the accumulated truncation errors act as a relevant perturbation steering the flow away from criticality.}. To verify that the implicitly scale invariant tensors are also explicitly approximately scale invariant after gauge fixing, we have constructed the radial transfer matrix MPO \Eq{eq:radialmpo} and calculated its smallest scaling dimensions (see Table~\ref{tab:scaledims}). Together with the coarse-graining procedure described in \Fig{fig:construction}, \Eq{eq:radialmpo} can be used to study fusion of primary fields and to calculate the operator product expansion coefficients of the underlying CFT, as has previously only been done using MERA-TNR for the Ising model \cite{Evenbly2016}. More importantly, our results suggest that the characteristic information of the underlying CFT can also be obtained from the fixed point MPS tensors \Eq{eq:fpeqs}, which in our formalism act as transparent building blocks for \emph{both} the linear and radial transfer matrix MPOs.

\begin{table}[t]
\caption{\label{tab:scaledims} Smallest scaling dimensions extracted from the eigenvalues of the radial transfer matrix MPO \Eq{eq:radialmpo} for the critical Ising model (left) and spin ice (right).}
\begin{ruledtabular}
\begin{tabular}{cllc}
   exact & Ising TNR$_{+}(6)$ & exact & Spin ice TNR$_{+}(10)$ \\
   \hline
   0.125 & 0.125236 & 1/6 & 0.163117 \\
   1 & 0.999282 & 1/6 & 0.167204\\
   1.125 & 1.123883 & 2/3 & 0.659684\\
   1.125 & 1.123883 & 2/3 & 0.662008\\
   2 & 1.998575 & 1 & 0.997413\\
   2 & 1.992823 & 1 & 0.997286\\
   2 & 1.996882 & 7/6 & 1.163503\\
   2 & 1.994090 & 7/6 & 1.163503   
\end{tabular}
\end{ruledtabular}
\end{table}

\begin{figure}[b]
 \includegraphics[width=1.0\linewidth,keepaspectratio=true]{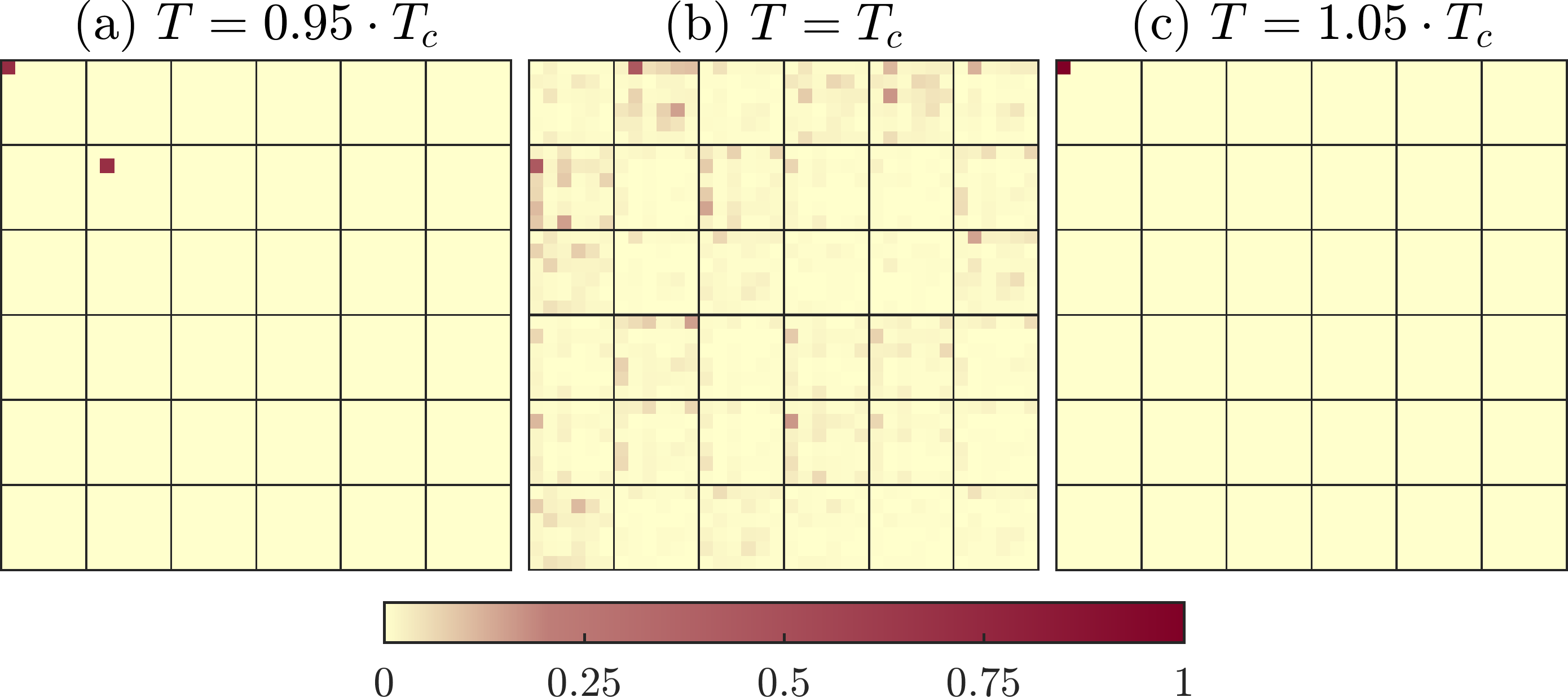}
 \caption{Nonnegative tensor elements of normalized fixed point tensors $C^{*}_{1,A}$ obtained from $D=6$ TNR$_{+}$ simulations of the Ising model at (a) $T<T_c$, (b) $T=T_c$, and (c) $T>T_c$. \label{fig:fptensors}}
\end{figure}

\simplesection{Effective Hamiltonians}
In \Fig{fig:fptensors}, we have plotted non-negative fixed point tensors \footnote{We have only plotted $C^{*}_{1,A}$, but the behavior of the other tensors $C^{*}_{2,A}$, $C^{*}_{1,B}$, $C^{*}_{2,B}$ is very similar. See also Appendix C.} for the Ising model at $T<T_{c}$, $T=T_{c}$, and $T>T_{c}$. Due to the element-wise nonnegativity, it is possible to equivalently consider the element-wise logarithm, so that we can interpret the tensor elements as energies of the configurations of the bonds connected to the site. The trivial tensor $C^{\mathrm{triv}}$ for $T>T_{c}$ has one dominant element, and all other arbitrarily small elements can be regarded as penalty terms in the effective Hamiltonian, signifying the use of a superfluous bond dimension in the description of the numerical fixed point. Similarly, for $T<T_{c}$, the $Z_{2}$ symmetry breaking tensor $C^{\mathrm Z_2}=C^{\mathrm{triv}} \oplus C^{\mathrm{triv}}$ is given by two equal dominant values with all other elements effectively zero. Both of these fixed points satisfy the algebraic relations \Eq{eq:fpeqs} since they are exact fixed points of the RG flow. Off-criticality we thus recover the fixed points previously found by Gu and Wen \cite{Gu2009}. The critical fixed point tensor for $T=T_{c}$ however is highly non-trivial, implying that the MPS optimization explores the full parameter space to approximate the exact fixed point which has infinite bond dimension. Due to the lattice geometry and the choice of the local coarse-graining transformation, the effective Hamiltonian encoded in the critical fixed point is given by local interactions between at most four effective $D$-dimensional degrees of freedom \footnote{A comprehensive analysis of the nature of these effective Hamiltonians will be reported elsewhere \cite{balunpub}.}\footnote{Although one might be tempted to extend the domain of the coarse-graining operation to even bigger blocks, there is of course a numerical trade-off between the locality of a coarse-graining scheme and the bond dimension that can be attained in practice.}. Note that the MPS tensors encoded in the critical fixed point, part of which is shown in \Fig{fig:fptensors}(b), provide an explicit and non-trivial example of numerically optimized solutions which \emph{approximately} satisfy the algebraic fixed point equations \Eq{eq:fpeqs} of the TNR$_{+}$ flow.

\simplesection{Conclusion and Outlook}
We have proposed a manifestly nonnegative tensor network renormalization algorithm to coarse-grain classical partition functions in real space, and provided additional evidence that tensor network renormalization techniques provide an approximation that behaves in a controlled way, introducing the required freedom to approximate the relevant physics at larger length-scales using effective interactions among effective degrees of freedom that are determined variationally. By restricting to nonnegative tensors, our work provides a bridge between heuristic block-spin prescriptions and modern tensor network approaches to coarse-graining.

Further improvement of the numerical results should be possible by taking lattice and internal symmetries into account and by improving the control on the gauge freedom. Due to the algorithm's formulation in terms of periodic MPS, we expect that the interplay with well-established theoretical and numerical MPS and MPO results will be of great importance in this regard. A generalization of our scheme to the quantum case is possible by constructing sequences of completely positive maps acting on projected-entangled pair states (PEPS) wave functions \cite{Verstraete:2004}. Another application would be to incorporate the formalism of MPO algebras \cite{Bultinck2017} in order to put topological restrictions on the CFT data extracted from tensor network renormalization \cite{Aasen2016,Hauru2016}.

\begin{acknowledgments}
\simplesection{Acknowledgements}
M.B.~would like to thank L.~Vanderstraeten, D.~Williamson and S.~Yang for discussions. This work is supported by an Odysseus grant from the FWO, a PostDoc grant from the FWO (J.H.), ERC grants QUTE and ERQUAF, and the EU grant SIQS.
\end{acknowledgments}

\bibliography{posrg.bib}

\onecolumngrid
\appendix
\clearpage

\section{Nonnegative matrix factorization\label{app:nmf}}
\subsection{Statement of the problem}
A nonnegative matrix $A \geq 0$ is a matrix in which all elements are equal to or greater than zero. Given a nonnegative matrix $A \in \mathbb{R}_{+}^{m \times n}$ and a factorization rank $k$, the problem of \emph{nonnegative matrix factorization} (NMF) is then to find a matrix decomposition $A \approx XY$, where $X \in \mathbb{R}_{+}^{m \times k}$ and $Y \in \mathbb{R}_{+}^{k \times n}$ are both nonnegative matrices as well. We can reformulate this problem as the following optimization problem:
\begin{align}
	\mathrm{argmin}_{X,Y}\nrm{A-XY}^2_{F}, \qquad X \geq 0,\quad Y \geq 0,\label{eq:costfuncnmf}
\end{align}
where $\nrm{\cdot}_{F}$ denotes the Frobenius norm. Note that, without the nonnegativity constraints, the optimal solution to \Eq{eq:costfuncnmf} is obtained via the singular value decomposition (SVD) of $A$.

It is clear that NMF is not unique in general because we can always insert a matrix $G$ and its inverse $G^{-1}$ such that the matrix product remains invariant,
\begin{align}
	XY=X G G^{-1} Y=\tilde{X}\tilde{Y}. \label{eq:gaugenmf}
\end{align}
If the two matrices $\tilde{X}=X G$ and $\tilde{Y}=G^{-1} Y$ are again nonnegative, they represent an equivalent parametrization $(\tilde{X},\tilde{Y})$ of the same factorization (X,Y). The requirements $\tilde{X}\geq 0$ and $\tilde{Y}\geq 0$ are surely satisfied if $G$ is a nonnegative monomial matrix $G=PD$, where $P$ is a permutation matrix and $D$ an invertible diagonal matrix containing only positive diagonal elements. More generally, there might also exist equivalent parametrizations $(XG,G^{-1}Y)$ with $XG \geq 0$ and $G^{-1}Y \geq 0$ where $G$ is not a monomial matrix, which can potentially spoil the uniqueness in a more severe way. Note that the smallest possible rank $k$ for which an \emph{exact} factorization $A=XY$ exists is the nonnegative rank of $A$, denoted with $\text{rank}_{+} (A)$. It satisfies $\text{rank} (A) \leq \text{rank}_{+} (A) \leq \min(m,n)$, and is defined as the smallest number of nonnegative vectors such that every column of $A$ can be written as a nonnegative linear combination of those vectors. When assuming $\text{rank} (A)=\text{rank}_{+} (A)=\text{rank} (X)=k$, a given exact factorization $(X,Y)$ of $A$ can be said to be unique if $A=\tilde{X}\tilde{Y}$ implies $\tilde{X}=XPD$ and $\tilde{Y}=(PD)^{-1}Y$, i.e.~if the only ambiguity of the factorization can be completely captured in terms of permutation and scaling matrices as defined above \cite{Gillis2012}.

\subsection{Intuition}
To understand why NMF is popular in the machine learning community, assume for instance that each $m$-dimensional column vector $A_{i}$ of $A \in \mathbb{R}_{+}^{m \times n}$ contains an element of a set of data. Finding $X \in \mathbb{R}_{+}^{m \times k}$ and $Y \in \mathbb{R}_{+}^{k \times n}$ so that $XY$ approximates $A$ as accurately as possible then corresponds to extracting $k$ features that capture latent properties of the dataset. Indeed, given the nonnegativity constraints, each element of the dataset is approximately reconstructed by summing over the $k$ basis elements in the columns of $X$ with coefficients given by the columns of $Y$, yielding a representation of the data which is a sum of distinctive parts. Applications include, but are not limited to, image processing, facial feature extraction, text mining and document classification, bioinformatics, recommender systems, clustering problems and spectral data analysis. Note however that, in general, the lack of uniqueness alluded to in \Eq{eq:gaugenmf} can be troublesome when the goal is to actually attribute significance to these emerging basis elements. For this reason, the uniqueness of NMF is closely linked to whether the numerically found features are really the only sensible interpretation of the data \cite{Donoho2004,Huang2014}.

\subsection{Algorithms}
In practice, the optimization problem \Eq{eq:costfuncnmf} has been shown to be NP-hard \cite{Vavasis2010}, and all available algorithms are only guaranteed to converge to a local optimum. The algorithm that kickstarted NMF developments was Lee and Seung's multiplicative update rule \cite{Seung1999}, 
\begin{align}
X \gets X \odot ((AY^{T})\oslash (XYY^{T})),\\
Y \gets Y \odot ((X^{T}A)\oslash (X^{T}XY)),
\end{align}
where $\odot$ and $\oslash$ denote Hadamard product and division respectively. It is an extremely simple alternating algorithm that updates the matrices element-wise, but has a rather slow convergence rate. Numerous variations and extensions have since been developed, and we refer the interested reader to Refs.~\cite{Wang2012,Gillis2014}. In practice, we  supplemented these algorithms by implementing a projected conjugate gradient approach (see Algorithm~\ref{alg:pcgnmf}) to improve solutions or convergence if required. Note that there is no agreed upon convergence criterion for NMF optimization, so in practice one is free to implement a strategy that takes into account cost function values, gradient norms, projected gradient norms, local tolerances, global tolerances, and maximum number of iterations.

\begin{algorithm}[H]
\caption{Projected conjugate gradient algorithm for NMF\label{alg:pcgnmf}}
\begin{algorithmic}[1]
\Procedure{PCGNMF}{$A$, $r$, \ldots} \Comment{Input a nonnegative matrix $A$ and a target rank $r$ (and convergence tolerances)}
\State $X_0,Y_0 \gets \Call{nndsvd}{$A, r$}$ \Comment{Initialize $X_0$ and $Y_0$}
\While{true} \Comment{Repeat until globally converged}
	\State $G_{0} \gets X_{0}Y_{0}Y_{0}^{T}-AY_{0}^{T}$, $D_{0} \gets -G_{0}$
	\While{true} \Comment{Repeat until locally converged for X}
		\State $\alpha \gets \Call{linesearch}{\ldots}$\Comment{Line search}
		\State $X_{1}=\max (0,X_{0}+\alpha D_{0})$ \Comment{Take step by projecting out all negative values}
		\State $G_{1}=X_{1}Y_{0}Y_{0}^{T}-AY_{0}^{T}$ \Comment{Compute new gradient}
		\State $\beta^{FR} \gets \nrm{G_{1}}^2/\nrm{G_{0}}^2$ \Comment{Calculate $\beta$ using your favourite formula (e.g.~Fletcher-Reeves)}
		\State $D_{1} \gets -G_{1}+\beta^{FR} D_{0}$ \Comment{Update search direction (with $\beta^{FR}=0$ for first iteration)}
		\State $D_{0} \gets D_{1}$, $G_{0} \gets G_{1}$, $X_{0} \gets X_{1}$\Comment{Prepare for next iteration}
		\If{\Call{islocalconverged}{$A, X_{0}, Y_{0}, \ldots$}}\Comment{Check local convergence (or maximum number of iterations)}
			\State \textbf{break}
		\EndIf
	\EndWhile
	\State $G_{0} \gets X_{0}^{T}X_{0}Y_{0}-X_{0}^{T}A$, $D_{0} \gets -G_{0}$
	\While{true} \Comment{Repeat until locally converged for Y}
		\State $\alpha \gets \Call{linesearch}{\ldots}$\Comment{Line search}
		\State $Y_{1}=\max (0,Y_{0}+\alpha D_{0})$ \Comment{Take step by projecting out all negative values}
		\State $G_{1}=X_{0}^{T}X_{0}Y_{1}-X_{0}^{T}A$ \Comment{Compute new gradient}
		\State $\beta^{FR} \gets \nrm{G_{1}}^2/\nrm{G_{0}}^2$ \Comment{Calculate $\beta$ using your favourite formula (e.g.~Fletcher-Reeves)}
		\State $D_{1} \gets -G_{1}+\beta^{FR} D_{0}$ \Comment{Update search direction (with $\beta^{FR}=0$ for first iteration)}
		\State $D_{0} \gets D_{1}$, $G_{0} \gets G_{1}$, $Y_{0} \gets Y_{1}$\Comment{Prepare for next iteration}
		\If{\Call{islocalconverged}{$A, X_{0}, Y_{0}, \ldots$}}\Comment{Check local convergence (or maximum number of iterations)}
			\State \textbf{break}
		\EndIf
	\EndWhile
	
	\If{\Call{isconverged}{$A, X, Y, \ldots$}}\Comment{Check convergence of $X$ and $Y$ together (or maximum number of iterations)}
		\State \textbf{break}
    \Else \Comment{Update local tolerances based on previous local tolerances, cost function values, and gradient norms}
        \State $\ldots \gets \Call{updatetolerances}{\ldots}$       
    \EndIf
\EndWhile
\State \textbf{return} $X_{1},Y_{1}$
\EndProcedure
\end{algorithmic}
\end{algorithm}

Another important aspect of NMF optimization is the choice of initialization $(X_0,Y_0)$. Starting from random nonnegative matrices surely is an option, but as NMF algorithms are local minimization algorithms, the choice of initial conditions can be crucial to the quality of the resulting local minimum and the speed of convergence. For our purposes, we favoured a semi-deterministic \texttt{NNDSVD} initialization based on the best rank-$k$ approximation of $A$ given by the SVD \cite{Boutsidis2008}. The initialization works by first calculating the subset of the $k$ largest singular values and vectors of $A$, i.e.~$U \Sigma V^{T}=\sum_{i=1}^{k} u_{i} \otimes v_{i}^{T}$, where the $k$ singular values $\Sigma$ appear in descending order and have been absorbed in the vectors $u_{i}$ and $v_{i}^{T}$. Each rank-one term $u_{i} \otimes v_{i}^{T}$ generally contains positive and negative values (apart from the dominant term due to Perron-Frobenius if the largest singular value is non-degenerate). A sensible nonnegative initialization is then obtained by replacing each $u_{i}\otimes v_{i}^{T}$, for $i=1,\ldots, k$, with the nonnegative outer products $\max (0,u_{i})\otimes \max (0,v_{i}^{T})$ or $\max (0,-u_{i})\otimes \max (0,-v_{i}^{T})$, depending on whichever has larger norm. The zero elements can be filled with small random values if need be.

Additionally, it can be convenient to fix the monomial gauge freedom \Eq{eq:gaugenmf}. This can be done by calculating the row vector $d_{X}$ containing the sum of each column of $X$ and column vector $d_{Y}$ containing the sum of each row of $Y$. We then insert the identity twice so that $XY=X\text{diag}(d_{X})^{-1}\text{diag}(d_{X})\text{diag}(d_{Y})\text{diag}(d_{Y})^{-1}Y$. After sorting the values on the diagonal of $\text{diag}(d_{X})\text{diag}(d_{Y})$ in descending order, we permute the columns of $X$ and the rows of $Y$ accordingly.

\section{Details on the TNR$_{+}$ implementation}
\subsection{Nonnegative tensor factorization\label{app:ntf}}
To coarse-grain a 2D bipartite lattice made up of rank-four tensors $[C_{1}]_{ijkl}\in \mathbb{R}_{+}^{d \times d \times d \times d}$ and $[C_{2}]_{ijkl}\in \mathbb{R}_{+}^{d \times d \times d \times d}$ in a manifestly nonnegative way, we will need to extend the nonnegative matrix factorization described above to nonnegative tensor factorization (NTF). Indeed, as pointed out in the main text and previously in Ref.~\cite{Yang2015}, we can interpret a block of four adjacent sites as a four-site periodic matrix product state (MPS) by reinterpreting $C_{1}$ and $C_{2}$ rank-three tensors after grouping the $d$-dimensional ``physical indices'' as follows
\begin{align}
	\vcenter{\hbox{ \includegraphics[width=0.15\linewidth]{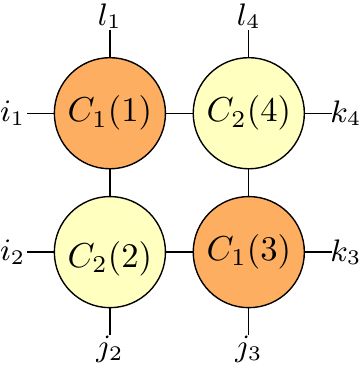} }} = \sum_{{\left\{\substack{i_1,i_2,j_2,j_3,\\ k_4,k_3,l_1,l_4}\right\}}} \tr \left( C_{1}(1)^{(l_{1}i_{1})} C_{2}(2)^{(i_{2}j_{2})} C_{1}(3)^{(j_{3}k_{3})} C_{2}(4)^{(k_{4}l_{4})} \right) \ket{l_{1}i_{1}}\ket{i_{2}j_{2}}\ket{j_{3}k_{3}}\ket{k_{4}l_{4}},
\end{align}
where the remaining $d$-dimensional indices have become ``virtual indices'', and are summed over in the matrix products. In the first step of the coarse-graining process, we construct an ansatz to approximate this block with a ``rotated block'' represented again by a ring of four sites with different rank-four tensors $[X_{n}]^{i_{n}j_{n}}_{\alpha_{n}\beta_{n}} \in \mathbb{R}_{+}^{D \times D \times d \times d}$, where $i_{n},j_{n}=1,\ldots,d$ (physical MPS dimension), and $\alpha_{n},\beta_{n}=1,\ldots,D \geq d$ (virtual MPS dimension). After grouping the physical dimension, we again obtain a periodic MPS representation of the block of sites. The cost function for the local approximation is then given by the constrained MPS overlap,
\begin{align}
	\mathrm{argmin}_{X_{1},X_{2},X_{3},X_{4}}\vcenter{\hbox{ \includegraphics[width=0.30\linewidth]{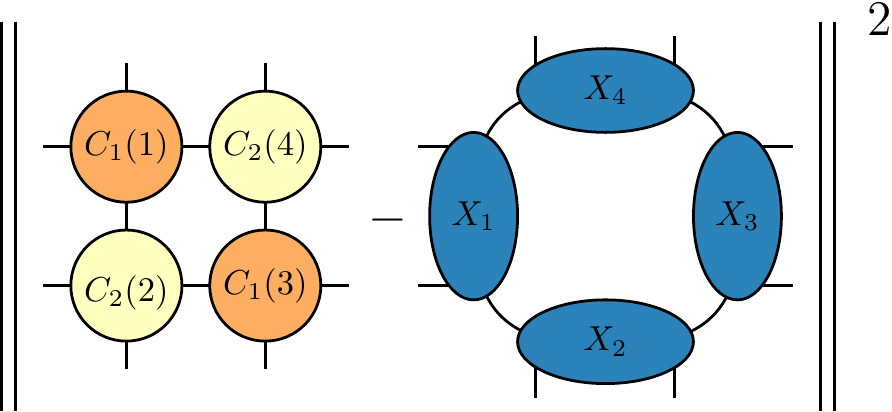} }}, \quad X_{1},X_{2},X_{3},X_{4} \geq 0, \label{eq:costfuncgraph}
\end{align}
or, after matching indices explicitly,
\begin{align}
\sum &\tr \left( C_{1}(1)^{(l_{1}i_{1})} C_{2}(2)^{(i_{2}j_{2})} C_{1}(3)^{(j_{3}k_{3})} C_{2}(4)^{(k_{4}l_{4})} \right) \ket{l_{1}i_{1}}\ket{i_{2}j_{2}}\ket{j_{3}k_{3}}\ket{k_{4}l_{4}} \\
&\approx \sum \tr \left( [X_{1}]^{(i_{1}i_{2})}[X_{2}]^{(j_{2}j_{3})}[X_{3}]^{(k_{3}k_{4})}[X_{4}]^{(l_{4}l_{1})}\right) \ket{i_{1}i_{2}}\ket{j_{2}j_{3}}\ket{k_{3}k_{4}}\ket{l_{4}l_{1}}. \label{app:costfunc}
\end{align}
Note that the original $C$-block and the ``rotated'' $X$-block have different tensor product structures. It is this breaking apart of the tensor product structure at each step which we believe to be a crucial feature of the success of all TRG-inspired methods.

\subsection{Sweeping projected conjugate gradient}
The cost function \Eq{eq:costfuncgraph} can be optimized using a generalization of the alternating inner loops of the \texttt{PCGNMF} procedure explained in Appendix~\ref{app:nmf} by reformulating the problem in terms of matrices. This is possible, as the cost function is equal to
\begin{align}
	\vcenter{\hbox{ \includegraphics[width=0.20\linewidth]{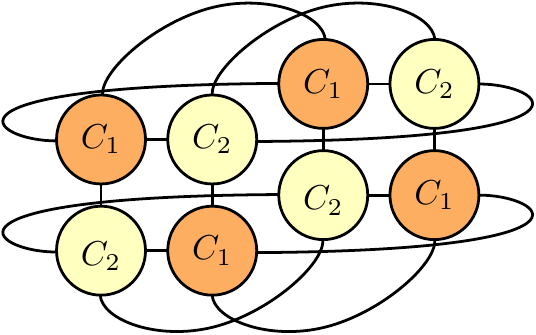} }} - 2 \vcenter{\hbox{ \includegraphics[width=0.17\linewidth]{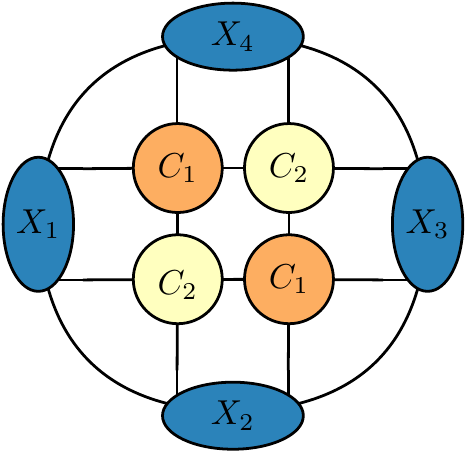} }} + \vcenter{\hbox{ \includegraphics[width=0.17\linewidth]{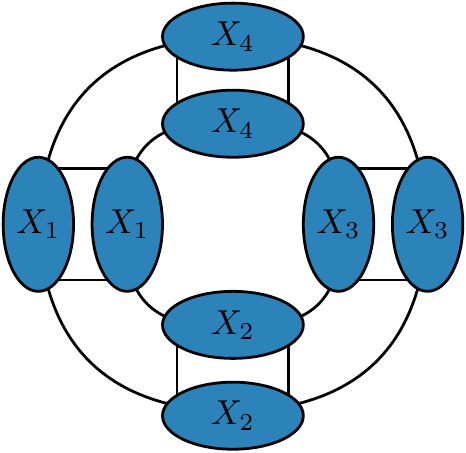} }}
\end{align}
Now assume we want to optimize $X_{1}$, keeping $X_{2}$, $X_{3}$, and $X_{4}$ fixed. Reshaping $X_{1}$ to a $D^2 \times d^2$ matrix and writing the gradient with respect to $X_{1}$ as the following $D^2 \times d^2$ matrix,
\begin{align}
	\text{grad}_{X_{1}} (x_{1}) = 2\vcenter{\hbox{ \includegraphics[width=0.14\linewidth]{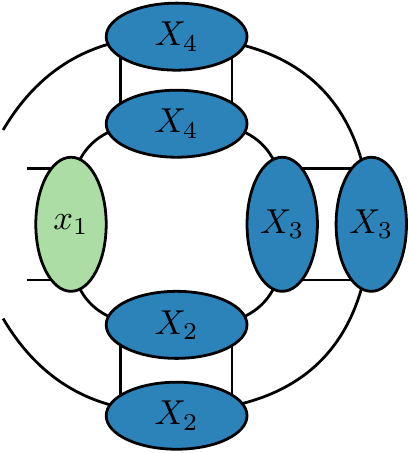} }} - 2 \vcenter{\hbox{ \includegraphics[width=0.14\linewidth]{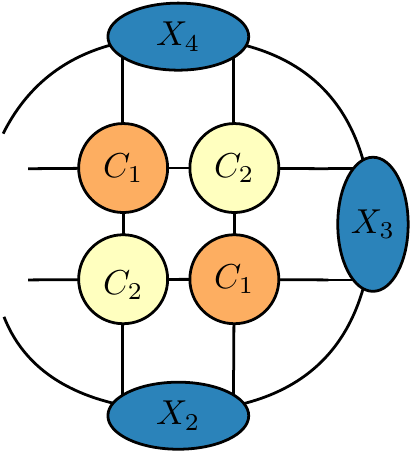} }},
\end{align}
we have all ingredients to implement an alternating projecting conjugate gradient method to sweep over all $X$ tensors.

One way to initialize the $X$ tensors is by constructing a tensor renormalization group (TRG) \cite{Levin2007} solution. If we use \texttt{PCGNMF} to find the rank-$D$ nonnegative decompositions of the $d^2 \times d^2$ matrices $C_{1}$ and $C_{2}$,
\begin{align}
	\vcenter{\hbox{ \includegraphics[width=0.20\linewidth]{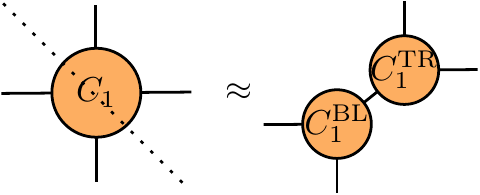} }}\quad \text{and} \quad \vcenter{\hbox{ \includegraphics[width=0.20\linewidth]{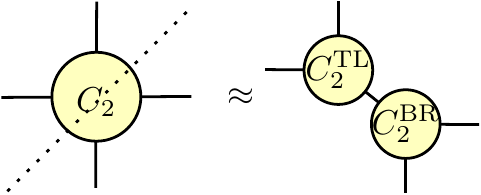} }},
\end{align}
our initilization looks like
\begin{align}
	\vcenter{\hbox{ \includegraphics[width=0.35\linewidth]{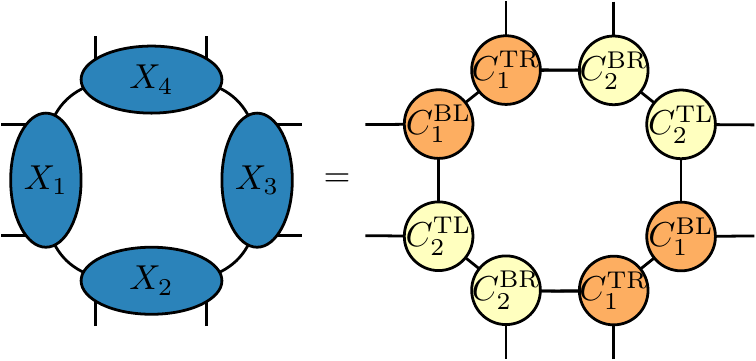} }}.
\end{align}

If we regard this particular initialization as a nonnegative TRG solution, we observe numerically that the local approximation error can be made significantly smaller than the initial solution. Just as MERA-TNR \cite{Evenbly2015b} and Loop-TNR \cite{Yang2015}, our TNR$_+$ algorithm is capable of systematically improving upon TRG. We also observed that the error does not keep on increasing at criticality, but remains approximately constant for a prolonged number of iterations. Off criticality, the error decreases quickly because the tensors flows to a simple fixed point encoding a trivial Hamiltonian.

\subsection{Coarse-graining\label{appcgconstr}}
From the optimized $X$ tensors, we can immediately construct the $C$ tensors of the coarse-grained lattice. For the general case under consideration where we impose neither lattice nor reflection symmetries, there is an ambiguity in constructing the new $C$ tensors of the next layer. We refer to Appendix~\ref{app:filtering} for an explanation as to why the ``plaquette'' grouping of the tensors mentioned in the main text is flawed. The ``vertex'' grouping, depicted below,
\begin{align}
	\vcenter{\hbox{ \includegraphics[width=0.50\linewidth]{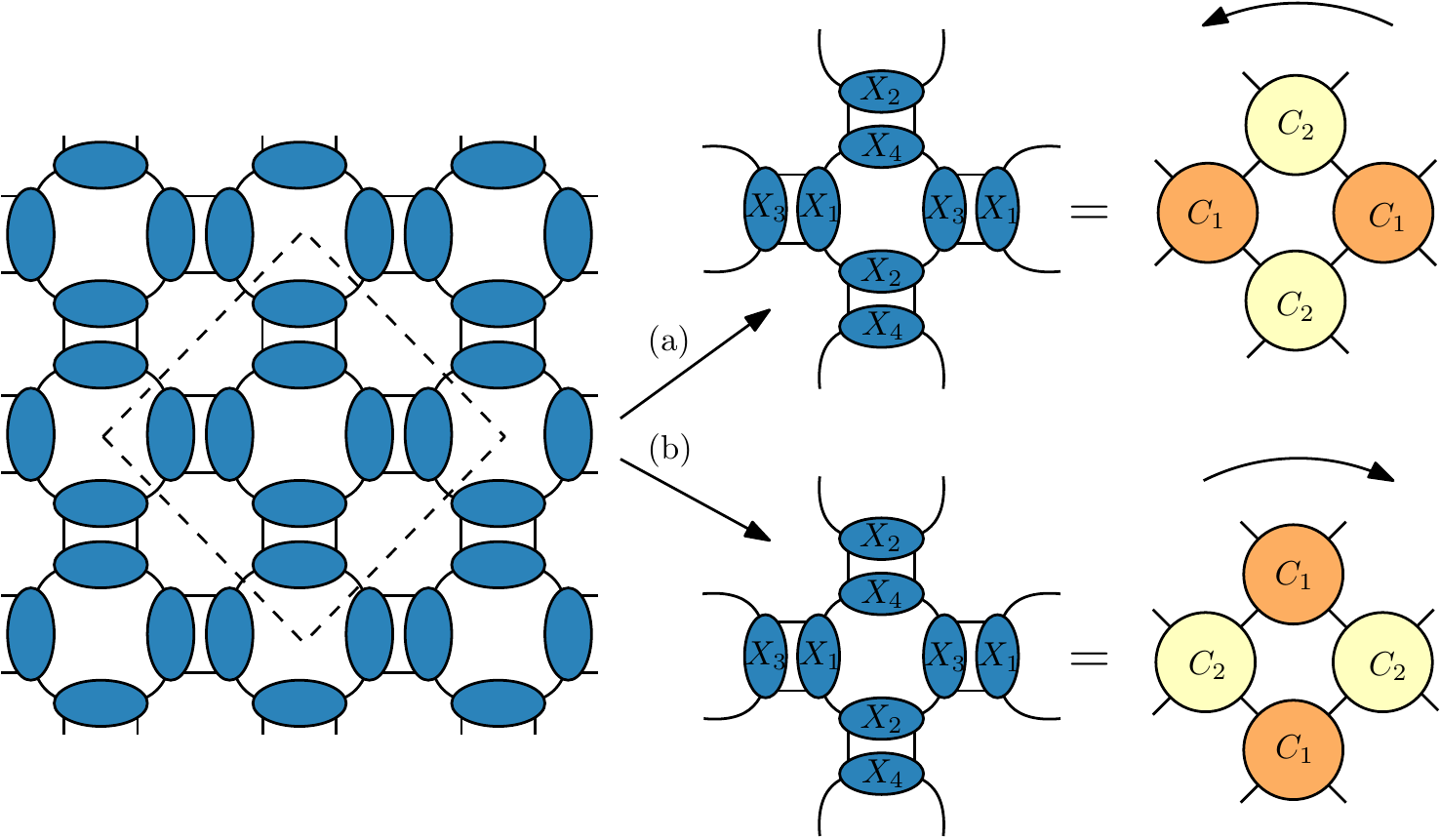} }}
\end{align}
can either be done by identifying the new $C$ tensors (and in this way choosing the orientation of the tilted lattice) counter-clockwise (a) clockwise (b), which is relevant for the construction of the radial transfer matrix MPO in Section~\ref{app:radtm}.

\subsection{A natural gauge choice for periodic nonnegative matrix product states\label{app:gauge}}
When implementing MPS optimization algorithms, choosing a canonical gauge is important both for manifestly revealing the entanglement content of a state as well as stabilizing the optimization through better conditioning of the matrices involved. Our TNR$_{+}$ algorithm can also benefit from fixing the nonnegative monomial gauge freedom \Eq{eq:gaugenmf} by providing a sensible basis for truncation purposes (see Appendix~\ref{app:filtering}), and by aiding in the recovery of explicit scale invariance (see Appendix~\ref{app:scaleinv}). We we will now describe a constructive way to fix the gauge freedom for a ring of nonnegative MPS tensors.

Without loss of generalization, consider a ring of four sites with rank-four tensors $[X_{n}]^{(ij)}_{(\alpha\beta)} \in \mathbb{R}_{+}^{D \times D \times d \times d}$, for $n=1,\ldots,4$, $i,j=1,\ldots,d$ (physical MPS dimension), and $\alpha,\beta=1,\ldots,D$ (virtual MPS dimension), so that the periodic nonnegative MPS is given by
\begin{align}
	\sum_{{\{i_nj_n\}}} \tr \left( X_{1}^{i_{1}j_{1}} X_{2}^{i_{2}j_{2}} X_{3}^{i_{3}j_{3}} X_{4}^{i_{4}j_{4}} \right) \ket{i_{1}j_{1}}\ket{i_{2}j_{2}}\ket{i_{3}j_{3}}\ket{i_{4}j_{4}} = \vcenter{\hbox{ \includegraphics[width=0.15\linewidth]{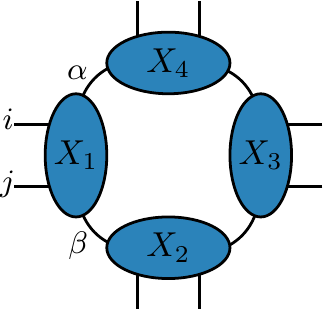} }}\label{fig:gauge_ring}
\end{align}
Let us now cut the bond connecting $X_{1}$ and $X_{4}$, and sum over all physical indices of the resulting tensor,
\begin{align*}
\vcenter{\hbox{ \includegraphics[width=0.40\linewidth]{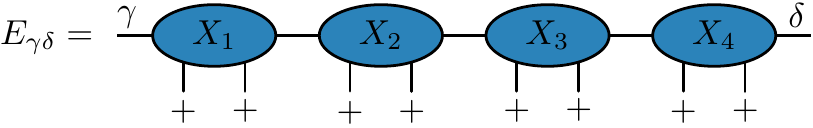} }},
\end{align*}
where $+$ denotes a vector of ones.
We then find the diagonal matrices $R$ and $C$ such that the matrix $M=REC$ becomes doubly stochastic, i.e.~$\sum_{i} M_{ij}=\sum_{j} M_{ij}=1$. Denoting the diagonals of $R$ and $C$ respectively as vectors $r$ and $c$, we can find these fixed point solutions by iterating
\begin{align*}
c = 1 ./ (A^{T}r), \quad r = 1./(Ac),
\end{align*}
which converges quickly as long as $E$ contains sufficiently many nonzero elements \cite{Knight2008}. We then substitute the identity twice on the bond that was cut, absorb $R$ and $C$ into $X_{1}$ and $X_{4}$ respectively, and obtain a central diagonal matrix,
\begin{align*}
\vcenter{\hbox{ \includegraphics[width=0.45\linewidth]{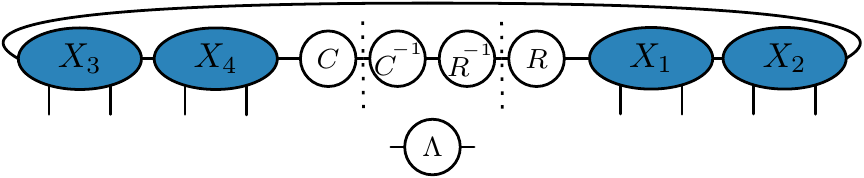} }},
\end{align*}
After sorting the diagonal elements of $\Lambda$ in descending order (which yields a permutation $P$), we can check for small values relative to the largest value and truncate up to some tolerance $\varepsilon$ (by means of an isometry $W$). In the end, we arrive at the following matrices to be absorbed into $X_{4}$ and $X_{1}$ respectively,
\begin{align}
G_{4}^{R}&=CP^{T}WW^{T}P\sqrt{\Lambda},\\
G_{1}^{L}&=\sqrt{\Lambda}P^{T}WW^{T}PR,
\end{align}
where the matrix $W$ is just the identity if there is no truncation or implicit truncation by setting the small singular values to zero, and an isometry onto the subspace that is retained if there is explicit truncation. Notice how $\Lambda$ gives us a nonnegative analogue of Schmidt values in the MPS case. The above gauge fixing can be repeated independently for all other bonds by permuting the tensors accordingly.

\section{Corner double line tensors and entanglement filtering\label{app:filtering}}
Corner double line tensors are a pathological case of non-critical fixed points of the TRG flow in the space of tensors. As argued in the main text, all TNR approaches (including the TEFR thanks to its entanglement filtering pre-processing step \cite{Gu2009}) are capable of removing CDL tensors because they surround a block of sites with a coarse-graining operation that can in principle detect correlations inside the block. Indeed, one can judiciously construct coarse-graining tensors that eliminate short-range correlations with a particular CDL structure \cite{Evenbly2015b}, which for TNR$_{+}$ proceeds as follows,
\begin{align}
\vcenter{\hbox{ \includegraphics[width=0.64\linewidth]{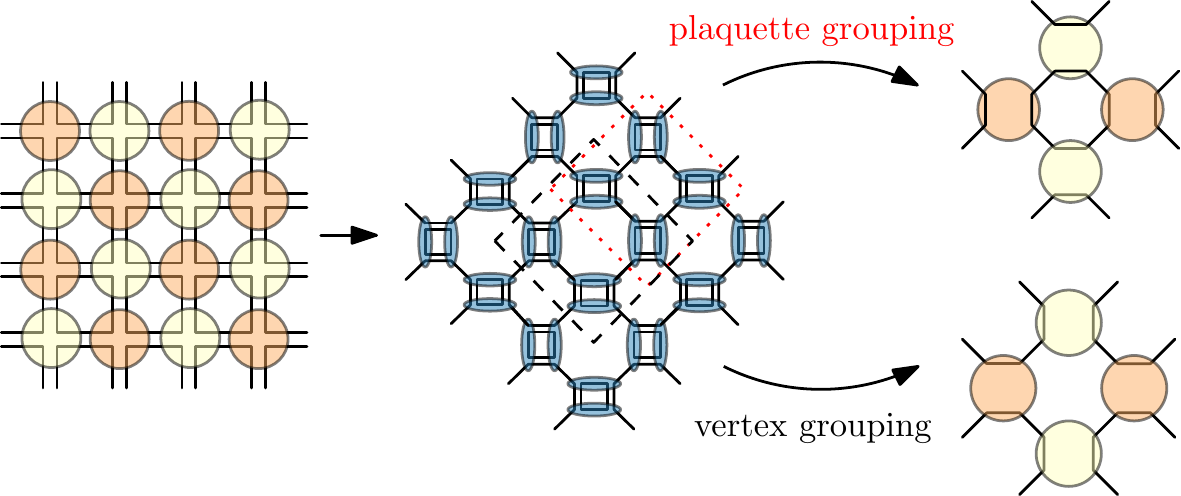} }},
\end{align}
The vertex grouping groups the plaquettes which \emph{do not} contain a loop, which results in a product state that can be approximated in the next iteration with a $D=1$ MPS. In contrast, grouping the plaquettes that contain loops reinstates corner double line tensors.

The above considerations however do not imply that numerical algorithms built on this premise will act accordingly, since CDL configurations are still local minima of the cost function and fixed points of the RG flow. It is important to mention that, in the ideal case above, the presence of CDL correlations is reflected in the degeneracies of the Schmidt values of the MPS. In practice however, there is no obvious way to detect these (approximate) tensor product structures inside the virtual bond and, numerically, there is often no structure to be inferred at all if local corner tensors contain non-degenerate eigenvalues. One way to deal with this is by monitoring the Schmidt values on the bonds of the ring \Eq{fig:gauge_ring} using the gauge fixing described in Appendix~\ref{app:gauge} to filter local correlations in a similar way to the tensor entanglement filtering step for TEFR and Loop-TNR discussed in Refs.~\cite{Gu2009,Yang2015}. Reformulated in conventional MPS language: what entanglement filtering does is truncating a periodic MPS (which here describes a block of sites of a classical 2D lattice model) by truncating its virtual dimension (which here contains short-range correlations ``inside" the block of sites). Note that a similar kind of truncation in MERA-TNR corresponds to alternating bond dimensions every step and inserting multiple optimized isometries at different stages in the actual implementation of the algorithm to reduce the intermediate bond dimensions and steer the optimization towards a preferred local minimum \cite{Evenbly2015a}. Another possible strategy, which requires no gauging, would be to maximize the overlap of a four-site periodic MPS containing irrelevant local details with a different MPS with a lower bond dimension and accept the lower-dimensional one if the fidelity is high enough. This can be done for nonnegative MPS with the algorithm described in Appendix~\ref{app:ntf}. Note that entanglement filtering has no effect for critical systems as there is in general no truncation possible due to the slowly decaying distribution of the Schmidt values (recall that the exact critical fixed point would require an infinite bond dimension), but by sorting diagonal values the permutations on the virtual indices can still be fixed.

\section{Symmetries and tensor network renormalization}
A straightforward application of Yang et~al.'s \cite{Yang2015} insight that it is worthwhile to model blocks of sites with periodic matrix product states (MPS), is the fundamental theorem of MPS \cite{Perez2008,Cirac2017} and its use in relating symmetries on the physical level to those on the virtual level. Consider an on-site symmetry operator acting on all sites, e.g.~ the flip operator
\begin{align}
	X=\begin{pmatrix} 
0 & 1 \\
1 & 0 
\end{pmatrix},
\end{align}
for the $Z_{2}$ symmetry of the Ising model. Invariance under the action of the symmetry implies that the transformed MPS should have an overlap with the original state that has modulus one. As such, the mixed transfer matrix of the original and the transformed state must have a dominant eigenvector with eigenvalue $|\lambda|=1$. It can then be shown that the effect of this statement is that the action of the symmetry on the physical level can be pushed through to the virtual level, up to a phase, which amounts to having a projective representation of the symmetry on the virtual level in the Schmidt basis. Now recall from the main text that the physical and virtual dimensions switch roles every RG iteration. This implies that the symmetry action on the virtual level becomes the action on the physical level of the next iteration, which in turn can be pushed through. In this way, the representation of the symmetry operator can be tracked throughout the entire coarse-graining network.

\section{Approximate scale invariance\label{app:scaleinv}}
As demonstrated in the main text, the TNR$_{+}$ algorithm yields tensors which correspond to approximate fixed points of the RG equations, and are approximately scale invariant at criticality. By observing gauge-invariant quantities, such as the eigenvalues of the linear transfer matrix (see Appendix~\ref{app:lintm}), it is clear that the fixed point tensors are implicitly approximately scale invariant and remain so for a large number of iterations. To recover explicit approximate scale invariance at the level of the individual tensor elements however, we need to fix the gauge freedom of the partition function across different scales. Note that the partition function written in terms of $C_{1}$ and $C_{2}$ remains invariant under the following transformations,
\begin{align}
	\vcenter{\hbox{ \includegraphics[width=0.27\linewidth]{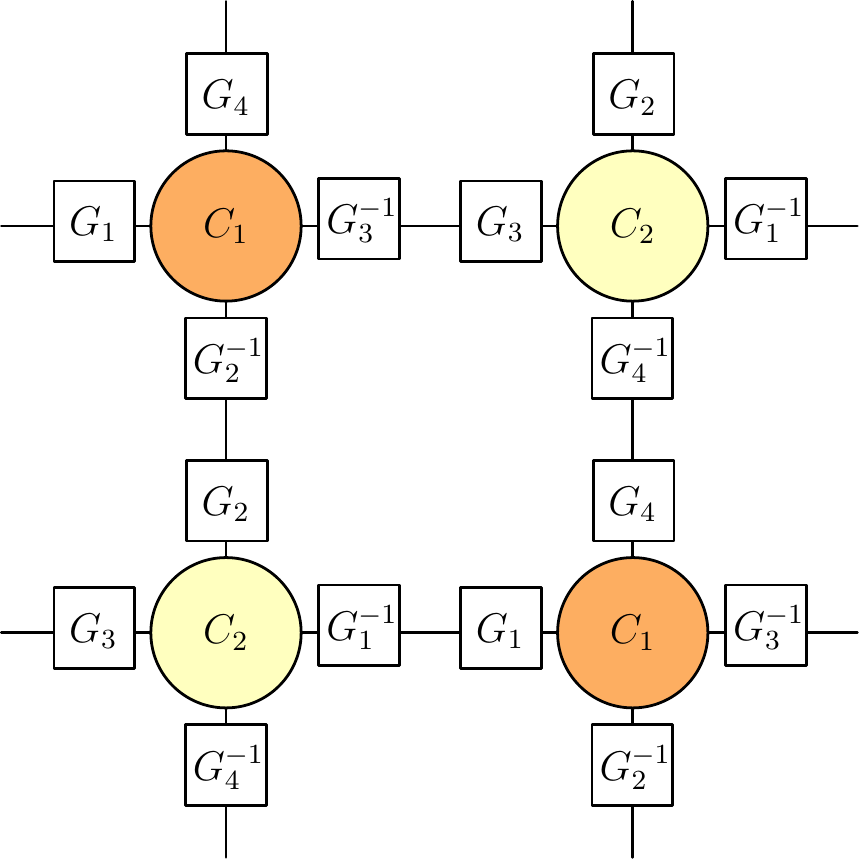} }}.\label{eq:gauges}
\end{align}
One simple way to achieve approximate scale invariance during the optimization itself is by adding additional constraints to the cost function \Eq{app:costfunc}. We can introduce a small penalty term $\lambda_{i}$ for each individual $X_{i}$ by adding
\begin{align}
	\sum_{i=1}^{4}\lambda_{i} \nrm{X^{(s-2)}_{i}-X^{(s)}_{i}}^2
\end{align}
to the cost function. This modified cost function will favor solutions which stay close to the previous equivalent solutions, i.e.~those of the even or odd iterations connecting lattices of the same orientation, which in turn renders the respective $C_{1}$ and $C_{2}$ tensors of the even and odd iterations approximately explicitly scale invariant as well.

\section{Conformal data from tensor networks\label{app:confdata}}
\subsection{Linear transfer matrix\label{app:lintm}}
Scaling dimensions of the conformal field theory (CFT) underlying a critical partition function can be extracted directly from its tensor network representation by constructing the linear transfer matrix \cite{Gu2009}. At every iteration step, we can construct the effective $2 \times 2$ and $4 \times 2$ row-to-row transfer matrices,
\begin{align}
	M^{(2)}=\vcenter{\hbox{ \includegraphics[width=0.12\linewidth]{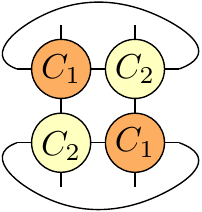} }}, \quad M^{(4)}=\vcenter{\hbox{ \includegraphics[width=0.21\linewidth]{lintm2} }},\label{eq:lintms}
\end{align}
whose gauge invariant eigenvalues can be directly related to the scaling dimensions of the primary operators and descendants of the CFT. The leading contribution to partition function (ignoring non-universal finite-size  corrections) on a torus of size $L_{x} \times L_{y}$ is given by
\begin{align}
	\mathcal{Z} \approx e^{aL_{x}L_{y}} \sum_{\alpha} e^{-2 \pi \frac{L_{y}}{L_{x}} (\Delta_{\alpha} - \frac{c}{12})},
\end{align}
where the non-universal contribution $e^{aL_{x}L_{y}}$ can be taken care of in the tensor network representation by properly normalizing the tensors, and $\Delta_{\alpha}$ and $c$ are respectively the scaling dimensions the central charge. We can then write the partition function $\mathcal{Z}=\tr (M^{L_{y}})$ in terms the row-to-row transfer matrix $M$, whose eigenvalue decomposition can be shown to be given by
\begin{align}
	M = \sum_{\alpha} e^{- \frac{2\pi}{L_{x}} (\Delta_{\alpha} - \frac{c}{12})} \ket{\alpha}\bra{\alpha},
\end{align}
which immediately yields numerical estimates for the scaling dimensions and the central charge given that $\Delta_{0}=0$. Note that we have assumed $M$ to be Hermitian, yet small deviations are to be expected numerically if no symmetries are enforced, resulting in distinct left and right eigenvectors.

As an aside, we can interpret the row-to-row transfer matrix of \Eq{eq:lintms} as an infinite MPO
\begin{align}
	\vcenter{\hbox{ \includegraphics[width=0.35\linewidth]{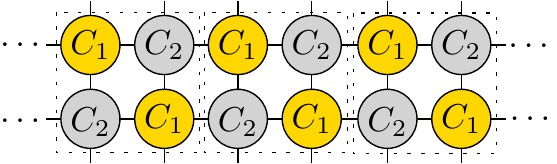} }}\label{eq:lintmmpo}
\end{align}
by blocking the tensors inside the dashed squares. Using the numerical MPO techniques recently developed in Ref.~\cite{Haegeman2016}, we can then calculate the low-lying excitation spectrum of this operator directly in the thermodynamic limit in terms of MPS excitation ans\"atze. As is to be expected, \Fig{fig:mpoexc} reveals a linear dispersion relation reflecting the continuum collapse of the CFT finite-size scaling results.

\begin{figure}[t]
 \includegraphics[width=0.35\linewidth,keepaspectratio=true]{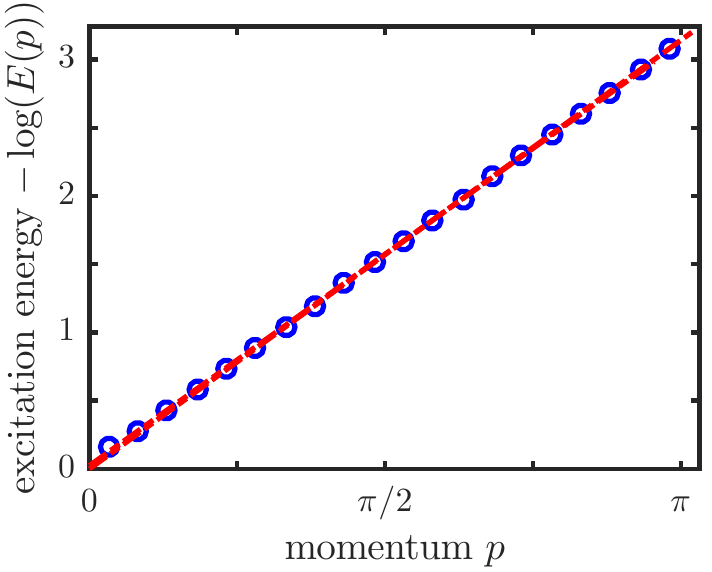}
 \caption{\label{fig:mpoexc}Linear energy-momentum spectrum of the critical Ising Hamiltonian encoded in the infinite row-to-row transfer matrix \Eq{eq:lintmmpo} obtained from a TNR$_{+}$ simulation with $D=6$ in the scale invariant regime. For the MPO fixed point calculations, a boundary MPS with bond dimension $\chi=18$ was used.}
\end{figure}

\subsection{Radial transfer matrix\label{app:radtm}}
Alternatively, we can extract conformal data from the radial transfer matrix, which can be obtained by tracking the RG flow around an open impurity \cite{Evenbly2016}. For the RG flow of TNR$_{+}$, we obtain after two iterations,
\begin{align}
	\vcenter{\hbox{ \includegraphics[width=0.50\linewidth]{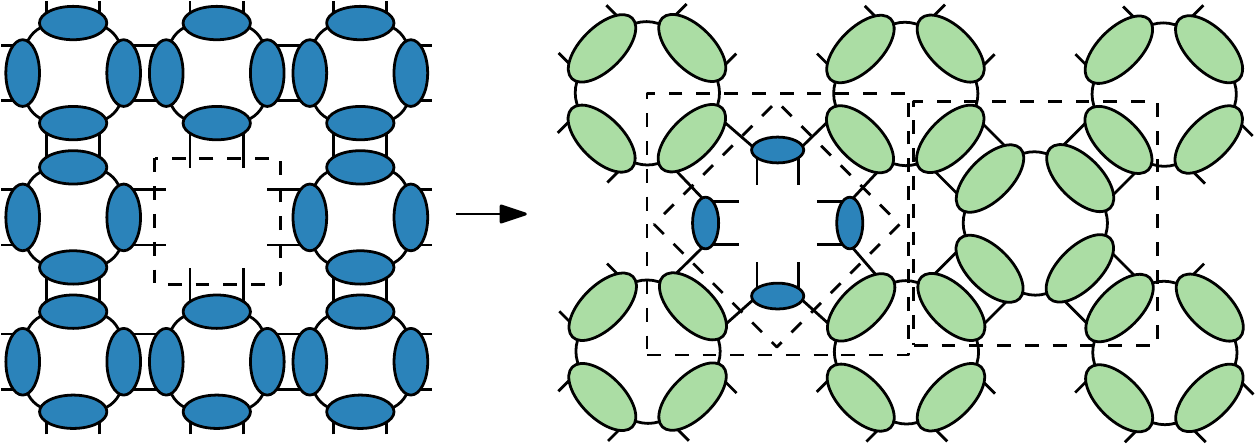} }},
\end{align}
where, for scale invariant systems, doing the next iteration everywhere on the block of tensors inside the rightmost bouding box gives rise to the same tensors as those obtained from the first iteration. Note that (see Appendix~\ref{appcgconstr}), we used different combinations of X-tensors in constructing the new C-tensors for even and odd iterations to arrive again at the same orientation after two iterations (``rotate back-and-forth''). If we would use the same contraction each iteration it would take eight iterations to again arrive at the original orientation (``rotate clockwise or counter-clockwise''), which would lead to a rather impractical superoperator. For scale invariant systems, we thus end up with repeated applications of the following MPO,
\begin{align}
	R=\vcenter{\hbox{ \includegraphics[width=0.15\linewidth]{radialtm} }},
\end{align}
which, after proper normalization, can be diagonalized to give
\begin{align}
	R=\sum_{\alpha} 2^{-\Delta_{\alpha}} \ket{\alpha}\bra{\alpha}.
\end{align}
Note that here again we have assumed $R$ to be Hermitian, yet small deviations are to be expected numerically if no symmetries are enforced, resulting in distinct left and right eigenvectors. If the gauge freedom has been fixed across RG steps, $\Delta_{\alpha}$ and $\ket{\alpha}$ are found to be respectively the scaling dimensions and approximate lattice representations of the primary fields and descendants. Indeed, for this construction to work, it is crucial to fix the gauge (see Appendix~\ref{app:scaleinv}), or else the degrees of freedom we deem equivalent do not match due to the different local gauges \Eq{eq:gauges}.

\end{document}